\documentclass[12pt,preprint]{aastex}
\shortauthors{PEREYRA \& MAGALH\~AES} 
\shorttitle{Polarimetry toward the {\it IRAS} Vela Shell} 

\begin{document} 
 
\title{Polarimetry toward the {\it IRAS} Vela Shell. II. Extinction and
 Magnetic Fields} 
 
\author{A. Pereyra and A. M. Magalh\~aes} 
\affil{Departamento de Astronomia, IAG, Universidade de S\~ao Paulo, Rua do
 Mat\~ao 1226, S\~ao Paulo, SP, 05508-900, Brazil} 
 
\email{antonio@astro.iag.usp.br} 
 
\begin{abstract} 
We explore correlations between visual extinction and polarization along the
 western side of the {\it IRAS} Vela Shell using a published polarimetric
 catalog of several hundreds of objects. Our extinction maps along this
 ionization front (I-front) find evidence of clumpy structure with typical
 masses between 1.5~and~6~{\it M}$_{\sun}$ and a mean length scale~{\it
 L}~$\sim$0.47~pc. The polarimetric data allowed us to investigate the
 distribution of the local magnetic field in small~($\sim$pc) scales across the
 I-front. Using the dispersion of polarization position angles, we find
 variations in the kinetic-to-magnetic energy density ratio of, at least, one
 order of magnitude along the I-front, with the magnetic pressure generally
 dominating over the turbulent motions. These findings suggest that the magnetic
 component has a significant contribution to the dynamical balance of this
 region. Along the I-front, the mean magnetic field projected on the sky is 
 [0.018~$\pm$~0.013]~mG. The polarization efficiency seems to change along the
 I-front. We attribute high polarization efficiencies in regions of relatively
 low extinction to an optimum degree of grain alignment. Analysis of the
 mass-to-magnetic flux ratio shows that this quantity is consistent with the
 subcritical regime ($\lambda < 1$), showing that magnetic support is indeed
 important in the region. Our data extend the overall $\lambda$$-${\it
 N}(H$_{2}$) relation toward lower density values and show that such trend
 continues smoothly toward low {\it N}(H$_{2}$) values. This provides general
 support for the evolution of initially subcritical clouds to an eventual
 supercritical stage. 
\end{abstract} 
 
 
\section{Introduction} 
\label{intro} 
 
Magnetic fields are important for the dynamics of expanding shells
 \citep{spi78,tro82,dra93}. Models of the interaction of shocks and ionization
 fronts in shells with magnetic fields exist in several configurations and media
 \citep{har97,pil90,pil94,red98,war98,wil00,wil01}. A proper determination of
 the magnetic field in shells is necessary to constraint these models.  
 
Following \citet[][hereinafter CF]{cha53}, the dispersion of polarization
 position angles, observed through a polarizing medium, permits to estimate the
 strength of the magnetic field component in the plane of the sky ({\it B}). In
 addition, this dispersion also yields information about the kinetic-to-magnetic
 energy density ratio \citep[$\rho $$_{kin}$/$ \rho $$_{mag}$,][]{zwe90}.  
 
Previous works used the CF procedure to calculate the magnetic field  projected
 on the sky in several regions of ISM \citep{gon90,mor92,chr94,ito99,hen01}.
 Nevertheless, a relatively large number of background stars are necessary in
 order to obtain a reasonable estimate for the strength of the magnetic field in
 a small area. Good statistics requires a sampling density higher enough to
 exceed that of the large-scale structures of the local magnetic field
 \citep{and05}.  
 
Polarimetric catalogues with several hundreds of objects in selected ISM regions
 and higher sampling density are now available for this type of study
 (\citealt{per02} - Paper I, \citeyear{per04}) with typical value of $\sim$1.2
 objects/arcminute$^{2}$ that is, at least, three orders of magnitude higher
 than previous works. 
 
The {\it IRAS} Vela Shell (IVS) encloses a cavity that appears to have been
 formed by stars of the Vela OB2 association through the effects of stellar
 winds and supernova explosions \citep{sa92}. However, the existence of the IVS
 is controversial and \citet{woe01} suggested that it is instead a density
 enhancement in the Gum nebula.  
 
A section of the western side of IVS (see Figures 2 and 3 in Paper I) shows a
 well-defined ionization front (I-front) seen almost edge-on with several
 magnetic field patterns. \citet{chu96} found that the kinetic energy is one
 order of magnitude higher than the gravitational potential energy in this
 region. They concluded that the structure is not gravitationally bound and
 would disperse on very short timescales were it not for the ram pressure of the
 expanding IVS, which continually sweeps up new interstellar matter into the
 cloud.

\clearpage 
\begin{figure} 
\includegraphics[scale=1]{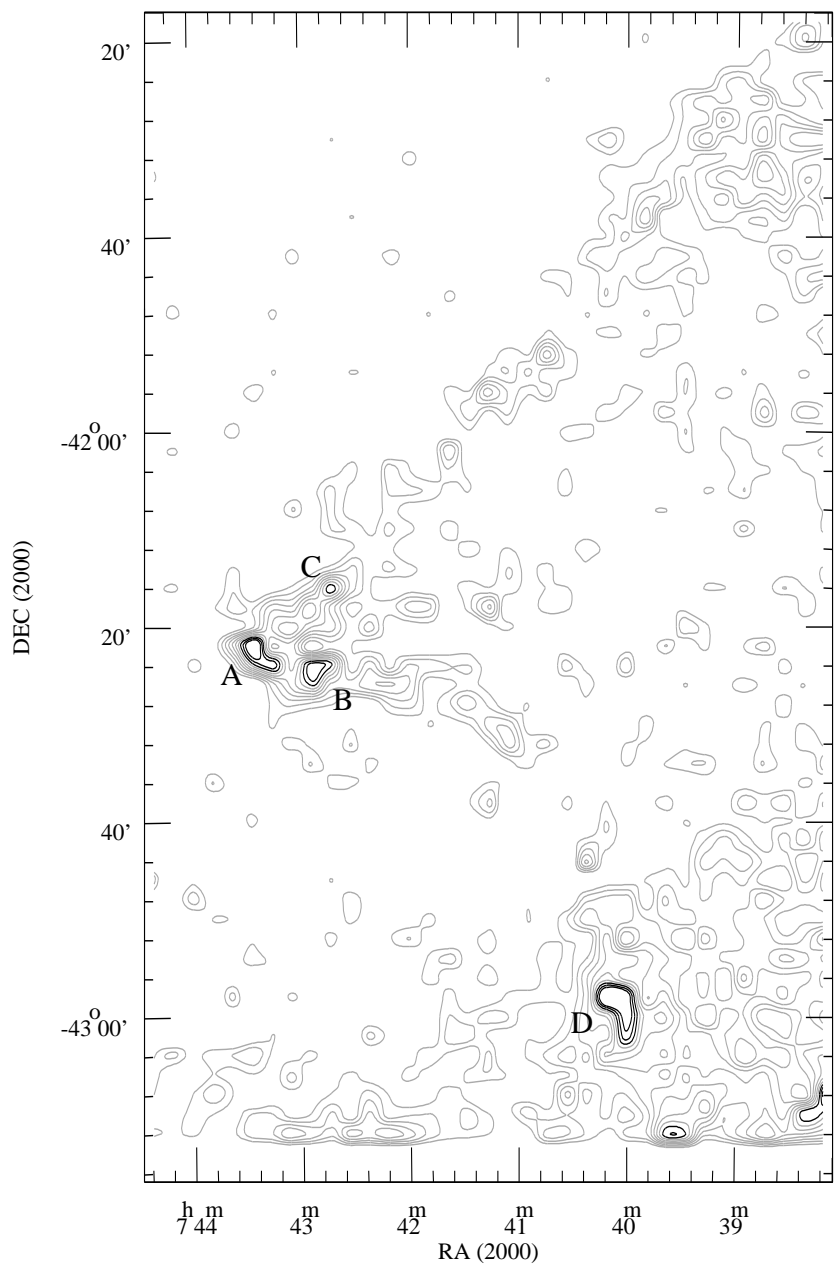} 
\caption{Extinction map toward the western side of {\it IRAS} Vela Shell. The
 gray contours go from 0.3 to 1.0 mag in steps of 0.1 mag. In black contours
 (from 1.0 to 1.2 mag) are the extinction structures (in letters) with {\it
 A}$_{\rm V}$ $\geq$ 1. \label{fcond}} 
\end{figure} 
\clearpage 
 
In this work we explore the role of the magnetic field on the dynamics of the
 IVS region using a published polarimetric catalogue (Paper I), with 856
 objects. In \S\ref{ext}, we describe the technique used to construct the
 extinction maps. We also detect extinction structures and estimate their
 masses. In \S\ref{stat}, we present a statistical analysis of the polarization
 catalogue. In \S\ref{polext}, we compare the polarization along the I-front
 with the detected extinction structures. In \S\ref{mag},  we apply the
 dispersion of polarization angle to infer the distribution of $\rho $$_{kin}$/$
 \rho $$_{mag}$ and the strength of {\it B} along the I-front. In \S\ref{polef},
 we explore the polarizing efficiency and its implications about the
 amplification of the local magnetic field. In \S\ref{masflux}, we analyze the
 magnetic support in terms of the mass-to-magnetic flux ratio. Conclusions are
 drawn in \S\ref{concl}. 
 
\clearpage 
\begin{deluxetable}{ccccccccc} 
\tablecolumns{9}  
\tablewidth{0pc}  
\tablecaption{Extinction structures.\label{tcond}}  
\tablehead{ 
\colhead{ID} & \colhead{RA$_{2000}$}  & \colhead{DEC$_{2000}$} & \colhead{{\it
 A$_{\rm V}$}} & \colhead{{\it N}(H$_{2}$)} & \colhead{area} & \colhead{{\it L}}
 & \colhead{mass} & \colhead{obs.}   \\ 
\colhead{} & \colhead{( h : m : s )} & \colhead{( $\degr$ : $\arcmin$ :
 $\arcsec$ )} & \colhead{(mag)} & \colhead{(10$^{21}$cm$^{-2}$)} & \colhead{(
 $\arcmin$$^{2}$ )}  & \colhead{(pc)} & \colhead{(M$\sun$)} & \\ 
\colhead{(1)} & \colhead{(2)} & \colhead{(3)} & \colhead{(4)} & \colhead{(5)} &
 \colhead{(6)} & \colhead{(7)} & \colhead{(8)} & \colhead{(9)}\\ } 
\startdata  
A & 07:43:26.90  & -42:22:22.55  & 1.42 & 1.33 & 8 & 0.52 & 3.73 &
 V12\tablenotemark{a}, A\tablenotemark{b} \\ 
B & 07:42:52.65 & -42:24:43.90  & 1.16 & 1.09 & 4 & 0.37 & 1.68 &
 B\tablenotemark{b} \\ 
C & 07:42:43.71 & -42:16:16.90  & 1.05 & 0.99 & 4 & 0.37 & 1.52 & -- \\ 
D & 07:40:05.85 & -42:58:11.84  & 1.50 & 1.41 & 12 & 0.64 & 5.95 & -- \\ 
\enddata 
\tablenotetext{a}{\citet{vil94}} 
\tablenotetext{b}{\citet{chu96}} 
\end{deluxetable} 
\clearpage 

\section{Visual Extinction by automatic star counts} 
\label{ext} 
 
\subsection{The extinction map} 
 
We used our own {\it IRAF}\footnote{{\it IRAF} is distributed by the National
 Optical Astronomy Observatory, which is operated by the Association of
 Universities for Research in Astronomy, Inc., under cooperative agreement with
 the National Science Foundation.} tasks to obtain maps of visual extinction of
 the region using an automatically star count technique. Firstly, we obtained
 positions of objects with stellar profiles from a  {\it Digitized Sky Survey}
 image (see Figure 3 in Paper I) using the {\it IRAF} task {\it daofind}. This
 routine automatically detects objects over certain threshold intensity. We used
 a threshold of 4 times of rms sky noise on the region. Our routines count the
 number of stars in each cell of 2$\arcmin \times$2$\arcmin$ defined on the {\it
 DSS} image and obtain the visual extinction according to \citet{dic78}: 
 
\begin{eqnarray*} 
{\it A_{\lambda }}   &  =      &    (1/{\it b}) log(   {\it n}_{0}   /   {\it n}_{1} )      \\ 
\end{eqnarray*}

where {\it n}$_{0}$ is the average number of stars in the comparison cell and
 {\it n}$_{1}$ is the total star number obtained by the count. We obtained star
 count values in regions away from the filamentary structure toward the
 north-east around ($\alpha$,$\delta$)$_{2000}$ = (07$^{\rm h}$44$^{\rm
 s}$,-41$\degr$27$\arcmin$) and computed an average {\it n}$_{0}$ = 62 for a box
 of 20$\arcmin\times$20$\arcmin$. \citet{van29} tables yields the {\it b}
 parameter and we used~{\it b}~=~0.44. This value is the best compromise from
 averaging several lines of sight toward darks clouds \citep{gre88,and96}. The
 maximum  error introduced by an incorrect {\it b} value is not more of 0.3 mag
 for {\it A}$_{\rm V}$ $ < $ 1.5 mag (see below).  
 
We transformed the extinction obtained from digitized photographic plates ({\it
 A$_{\rm phot}$}) to visual extinction {\it A$_{\rm V}$} ($\lambda$5500$\rm
 \AA$) using the normalize extinction curve from \citet{ble72}. The used {\it
 DSS} image corresponds to Schmidt plates with the emulsion/filter combination
 $\rm IIIaJ/GC395$ centered in 4500$\rm \AA$ \citep{las90}. We obtained the
 necessary correction by a polynomial fit in the range 0.4$-$3.4$\micron$ of the
 normalized extinction curve from \citet{ble72} and then recalculated the
 extinction into the photographic band. The resulting relation was: 
 
\begin{eqnarray*} 
    {\it A}_{\rm V}    &   =     &  0.79  {\it A}_{\rm phot}        \\ 
\end{eqnarray*} 
 
We then built an extinction image of 30$\times$60 cells (=~1$\degr
 \times$2$\degr$) for the region. The error in each cell grows with the
 extinction value  \citep{dic78}. For the cells with no star we obtained a 
 lower limit fitting the residuals between consecutive steps of extinction and
 extrapolated the correction for the case {\it n}$_{1}$ = 0. This procedure
 resulted in assuming {\it n}$_{1}$ = 0.28 for cells with no star and the
 corresponding lower limit of detectable maximum extinction was 4.2 mag.
 Nevertheless, in the regions of interest (i.e. where polarimetric data exist)
 the extinction is relatively low with maximum values of 1.5 mag ({\it A$_{\rm V
 max}$}), well below than that limit. Figure~\ref{fcond} shows the extinction
 maps obtained with this technique. The contours go from 0.3 to 1.2 mag in steps
 of 0.1 mag. 
 
We must note that procedure used here to estimate the extinction is differential
 in origin. It means that we computed the extiction with respect to the
 comparison cell {\it n}$_{0}$. In order to check if a  zero-point bias is
 present in our extinction maps, we used the reddening maps of \citet{sc98}. The
 average extinction in a radius of 5$\arcmin$ toward the region of the
 comparison cell was 1.06$\pm$0.04\footnote{see the Dust Extinction Service
 available at http://irsa.ipac.caltech.edu/applications/DUST/} mag. We consider
 that this level may represent the foreground extinction to the I-front.
 Therefore, we purposely did not correct our extinction maps by this level in
 order to include just the extinction associated with the I-front.  
 
\clearpage
\begin{deluxetable}{ccccccccccccccc} 
\tablecolumns{15} 
\tablewidth{0pc}
\rotate 
\tabletypesize{\scriptsize} 
\tablecaption{Polarization analysis.\label{tanl}} 
\tablehead{
\colhead{field/{\it N}$_{\rm cat}$\tablenotemark{a}}  & \colhead{trend}		       & \colhead{$\theta_{gauss}$}	          &
\colhead{$\sigma_{\theta gauss}$}	                    & \colhead{{\it N}}                & \colhead{$\langle P\rangle$}              &      \colhead{$\langle\sigma\rangle$}                      & \colhead{$\langle\theta\rangle$}	& \colhead{$\langle\sigma_{\theta}\rangle$} &      \colhead{$\Delta\theta$}		                    & \colhead{$(\Delta\theta)^{2}$}   & \colhead{$\langle A_{\rm V}\rangle$}      &  \colhead{$\langle\it n\rm (H_{2})\rangle$}                          & \colhead{{\it B}}                & \colhead{$\lambda$}   \\ 
\colhead{}            & \colhead{}       & \colhead{(deg)}   & \colhead{(deg)}    & \colhead{}                & \colhead{(\%)}    & 
\colhead{(\%)}        & \colhead{(deg)}  & \colhead{(deg)}   & \colhead{(deg)}    & \colhead{10$^{-2}$}       & \colhead{(mag)}   &
\colhead{(cm$^{-3}$)} & \colhead{(mG)}   & \colhead{} \\
\colhead{(1)}  & \colhead{(2)}    & \colhead{(3)}     & \colhead{(4)}   & \colhead{(5)}   & \colhead{(6)}   & \colhead{(7)}    & 
\colhead{(8)}  & \colhead{(9)}    & \colhead{(10)}   & \colhead{(11)} & \colhead{(12)} & \colhead{(13)} & \colhead{(14)} & \colhead{(15)}    \\ }
\startdata 
01/14 & 1 &  \nodata &  \nodata & 14 &  1.105 &  0.005  &  32.9  &   1.2  & \nodata & \nodata  & 0.34$\pm$0.24 & 280 & \nodata & \nodata \\
02/71 & 1 &  63.7    &  18.1    & 66 &  0.665 &  0.001  &  61.2  &   1.1  &  18.0   &  9.9     & 0.31$\pm$0.10 & 255 & 0.008   & 0.27 \\
03/68 & 1 &  69.8    &  15.2    & 43 &  0.703 &  0.005  &  57.9  &   1.8  &  15.1   &  6.9     & 0.23$\pm$0.12 & 189 & 0.009   & 0.19  \\
04/59 & 1 &  96.5    &  13.2    & 35 &  1.798 &  0.003  & 104.0  &   0.7  &  13.2   &  5.3     & 0.65$\pm$0.29 & 535 & 0.016   & 0.28 \\
      & 2 &  40.9    &  12.3    & 22 &  0.870 &  0.001  &  41.8  &   1.0  &  12.3   &  4.6     & 0.36$\pm$0.17 & 296 & 0.013   & 0.20 \\
05/76 & 1 &  40.2    &  5.4     & 74 &  2.284 &  0.001  &  42.4  &   1.0  &   5.3   &  0.9     & 0.42$\pm$0.18 & 346 & 0.033   & 0.09 \\
06/68 & 1 &  10.8    &  14.9    & 66 &  1.723 &  0.001  &  10.0  &   0.9  &  14.9   &  6.8     & 0.41$\pm$0.14 & 337 & 0.012   & 0.25 \\
07/61 & 1 &  171.6   &  10.2    & 51 &  2.529 &  0.003  & 165.5  &   0.8  &  10.2   &  3.1     & 0.38$\pm$0.10 & 313 & 0.016   & 0.17 \\
      & 2 &  23.5    &  2.9     &  8 &  2.344 &  0.012  &  23.2  &   0.7  &   2.8   &  0.2     & 0.45$\pm$0.16 & 370 & 0.064   & 0.05 \\
08/68 & 1 &  32.5    &  8.0     & 65 &  1.701 &  0.003  &  32.9  &   1.0  &   7.9   &  1.9     & 0.32$\pm$0.08 & 263 & 0.019   & 0.12 \\
09/18 & 1 & \nodata  & \nodata  & 18 &  0.429 &  0.005  &  95.9  &   1.7  & \nodata &  \nodata & 0.21$\pm$0.10 & 173 & \nodata & \nodata \\
10/78 & 1 & \nodata  & \nodata  & 78 &  0.057 &  0.001  &  47.3  &   1.1  & \nodata &  \nodata & 0.18$\pm$0.09 & 148 & \nodata & \nodata \\
11/55 & 1 & 105.3    &  9.5     & 43 &  0.598 &  0.002  &  109.4 &   1.2  &   9.5   &  2.7     & 0.33$\pm$0.12 & 271 & 0.016   & 0.15 \\
12/62 & 1 &  15.9    &  14.3    & 29 &  1.623 &  0.001  &  16.0  &   1.0  &  14.3   &  6.2     & 0.44$\pm$0.34 & 362 & 0.012  & 0.25 \\
      & 2 &  116.3   &  19.0    & 33 &  0.799 &  0.001  &  116.1 &   0.9  &  19.0   &  11.0    & 0.47$\pm$0.31 & 387 & 0.010  & 0.35 \\
13/50 & 1 &  104.6   &  10.1    & 44 &  3.113 &  0.002  &  104.2 &   1.0  &  10.1   &  3.1     & 0.42$\pm$0.16 & 346 & 0.017  & 0.17 \\
14/63 & 1 &  75.6    &   8.4    & 23 &  0.435 &  0.001  &  75.6  &   0.9  &   8.3   &  2.1     & 0.27$\pm$0.09 & 222 & 0.017  & 0.12 \\
      & 2 &  8.1     &  19.3    & 31 &  0.220 &  0.001  &  9.4   &   1.3  &  19.3   &  11.3    & 0.28$\pm$0.14 & 230 & 0.007  & 0.27 \\
      & 3 &  143.1   &   8.9    &  9 &  0.733 &  0.004  & 133.0  &   0.8  &   8.9   &  2.4     & 0.36$\pm$0.15 & 296 & 0.018  & 0.14 \\
15/12 & 1 & \nodata  & \nodata  & 12 &  0.218 &  0.003  & 158.4  &   0.9  & \nodata &  \nodata & 0.58$\pm$0.34 & 477 & \nodata & \nodata \\
16/52 & 1 & \nodata  & \nodata &  52 &  0.251 &  0.001  & 125.1  &   1.1  & \nodata &  \nodata & 0.51$\pm$0.12 & 420 & \nodata & \nodata \\
\enddata
\tablenotetext{a}{original number of objects in each field from catalog (Paper I).}
\end{deluxetable} 
\clearpage

 
\begin{figure*} 
\includegraphics[scale=1]{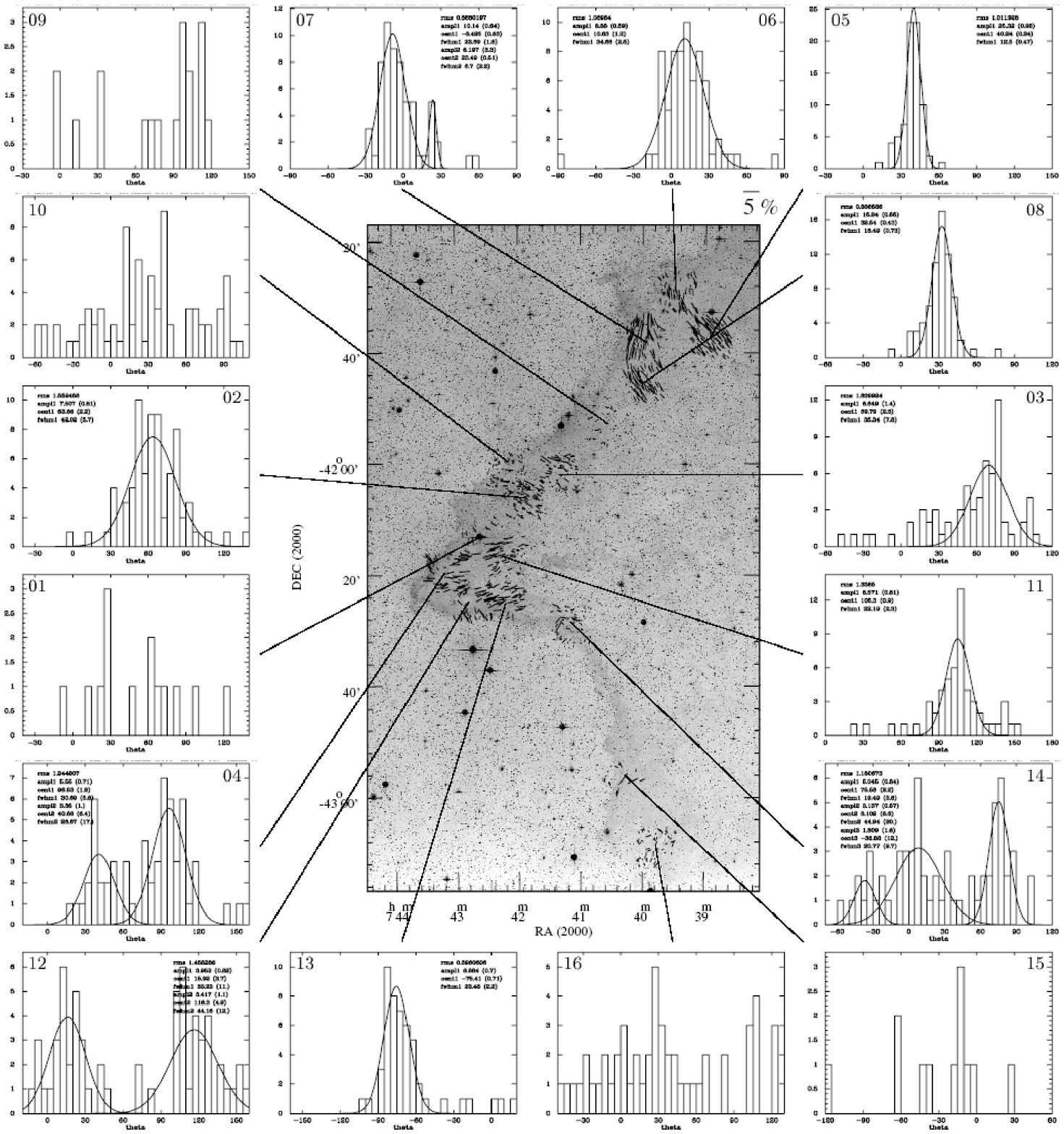} 
\caption{Polarization map for IVS region (see Paper I) with $ \theta $
 histograms for each position analyzed (in numbers). The gaussian fits used to
 obtain the dispersion of polarization angles ($ \Delta \theta $) in each case
 are also shown.\label{fhist}} 
\end{figure*} 
 
\clearpage 
 
\subsection{Analysis of extinction and  masses.} 
\label{extmas} 
 
As noted by \citet{chu96}, the IVS far-infrared dust emission follows the
 morphology of the I-front, implying that the radiation field of Vela OB2 and
 shocks induced by the expansion of the shell must heat the dust. These authors
 also found that the CS emission accompanies the I-front and the emission from
 dust.  
 
Analyzing Figure~\ref{fcond} we observe that the extinction structures accompany
 the I-front pattern and a good correlation with the extended emission of dust
 exists (see Figure 2 in Paper I). Therefore, it is evident that dust, which
 originates the observed extinction, was shifted to the west of the I-front but
 accompanying its eastern pattern including the north and south curvatures
 observed on it. Although the extinction values are relatively small (up to
 $\sim$1.5 mag), the morphology observed on the extinction map suggests a clumpy
 region also invocated by \citet{chu96} using CS maps.

In order to better visualize the extinction structures, we choose regions with
 {\it A$_{\rm V}$} $ \geq $ 1 mag. This level is a good compromise
 ($\sim$$\frac{2}{3}$ {\it A$_{\rm V max}$}) to detect significant structures in
 our maps. Figure~\ref{fcond} shows these structures. We analyse some physical
 properties of them in Table~\ref{tcond}. In columns (2) and (3) we show the
 equatorial coordinates for the positions with the highest extinction observed
 in each structure. In column (4) is the highest extinction value in each
 structure. Column (5) shows the molecular hydrogen column density estimated at
 the position of highest extinction in each structure using the standard
 gas-to-extinction ratio \citep[{\it N}$_{\rm H_{2}}$/{\it A}$_{\rm V}$ = 0.94
 $\times$ 10$^{21}$ cm$^{-2}$ mag$^{-1}$,][]{boh78}. This ratio assumes that
 most of hydrogen is in molecular form and {\it R$_{\rm V}$} $=$ 3.1. In column
 (6) is the area in square arcminutes associated with each structure and
 estimated by adding all the cells with {\it A}$_{\rm V}$ $ \geq $ 1 inside of
 it. Column (7) shows the dimension associated with the structure in parsecs and
 estimated by $ \sqrt{2\rm (area)}$. Finally, in column (8) is a lower limit of
 the mass of each structure estimated using \citet{dic78}: 
 
\begin{eqnarray*} 
M  & = & (\alpha d)^{2}(N_{\rm H_{2}}/A_{\rm V})m \sum_{i} A^{i}_{\rm V}, \\ 
\end{eqnarray*} 
 
where $ \alpha $ is the cell size in radians, {\it d} is the distance to the
 cloud in centimeters,  
({\it N}$_{\rm H_{2}}$/{\it A}$_{\rm V}$) is the standard gas-to-extinction
 ratio, ${\it m} = \mu_{\rm H_{2}} {\it m}_{H} =  2.8{\it m}_{H}$ is the mean
 particle mass (allowing for 10\% He by number), $\mu_{\rm H_{2}}$ is the mean
 molecular weight with respect to the number of H molecules, and ${\it A}{i
 \atop \rm V}$ is the visual extinction in each reseau element {\it i}.  
 
Among the four structures (A, B, C and D) shown in Figure~\ref{fcond}, only the
 A structure is coincident with the condensation named V12 in
 \citet[][hereinafter VMF]{vil94} in the Vela region. In the southern dark
 clouds catalogue of \citet{har86}, this structure has {\it A}$_{\rm V}$ = 6
 mag, and in VMF the extinction is {\it A}$_{\rm V}$~=~3.7 mag using C$^{18}$O
 linewidths. Our extinction value ({\it A}$_{\rm V}$~=~1.42~mag) is clearly the
 lowest. VMF also estimated the scale size, hydrogen column density and mass for
 V12 and obtained 0.15~pc, 3.4~$\times$~10$^{21} $cm$^{-2}$ and 1.3M$\sun$
 (lower limit), respectively. The differences with our values in
 Table~\ref{tcond} can be traced to a lower distance used by VMF (300 pc versus
 450 pc in this work) and the conversion factor of C$^{18}$O column density to
 visual extinction used by VMF  \citep[{\it A}$_{\rm V}$ = 6.4 $\times$
 10$^{-15}${\it N}(C$^{18}$O) + 3.2 mag from][]{noz91}. In contrast, we obtained
 our extinction values directly from the {\it DSS} image.
 
The A and B structures in our extinction map are coincident with A and B clumps
 seen in the distribution study of CS column density of \citet{chu96} done in
 the same region. In summary, from Table~\ref{tcond} we estimate a range for
 typical masses of 1.5$-$6 ${\it M}_{\sun}$ and a mean length scale {\it L}
 $\sim$0.47 pc for the extinction structures found by us. 
 
\clearpage 
 
\begin{figure} 
\includegraphics[scale=1]{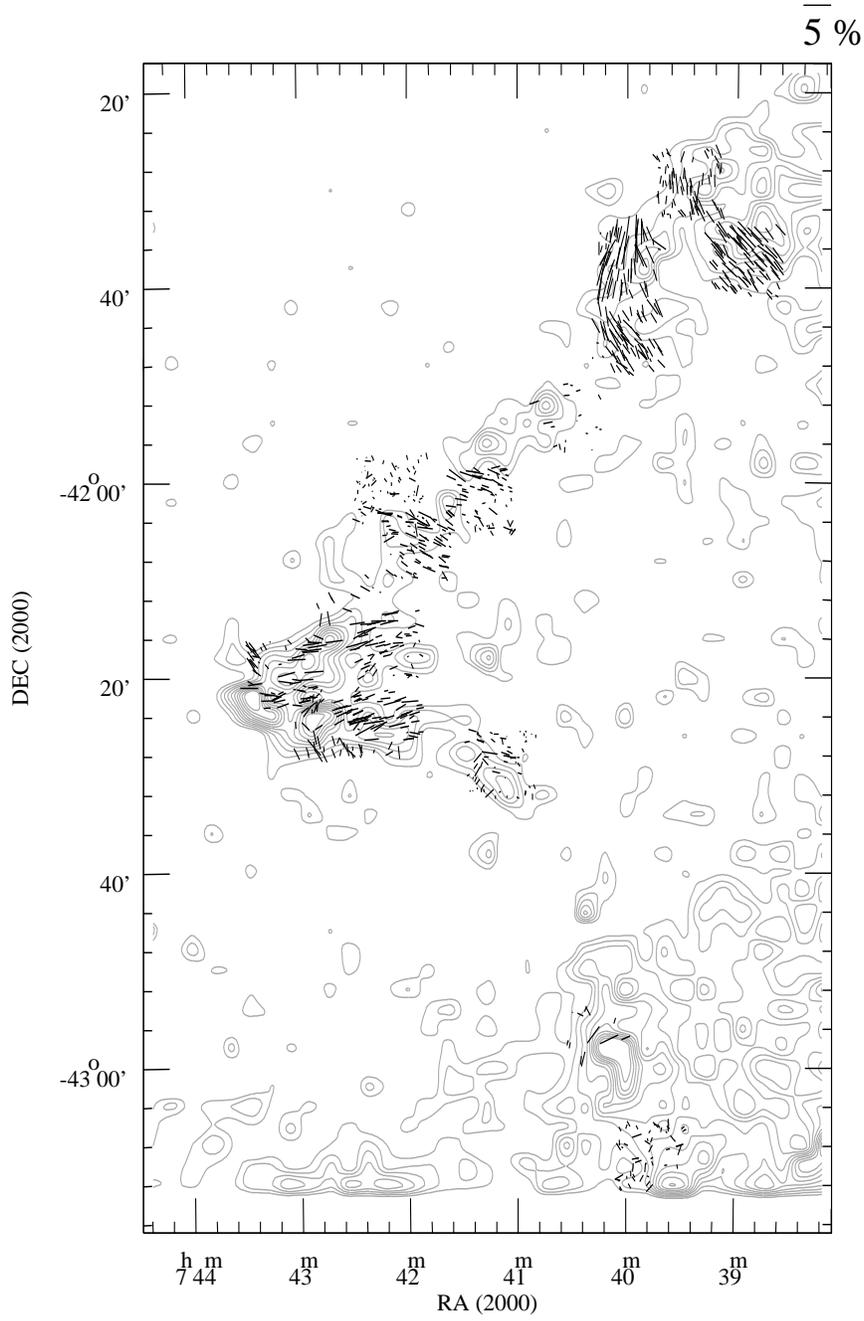} 
\caption{Polarization map overlaying the extinction map. The polarization scale
 is shown on the upper right. The contours are the same as in
 Fig.~\ref{fcond}.\label{fextpol}} 
\end{figure} 
 
\clearpage 
 
\section{Polarization analysis} 
\label{stat} 
 
In order to better quantify the observed polarization pattern in each one of
 sixteen fields studied in Paper I, we present a statistical analysis in
 Table~\ref{tanl}. Each field is given in column (1), together with the catalog
 stars within the field ({\it N}$_{\rm cat}$). We construct polarization angle
 histograms (see Figure~\ref{fhist}) for each field and used gaussian fits to
 obtain the representative polarization angle $\theta_{gauss}$ and its
 dispersion $\sigma_{\theta gauss}$, indicated in colums (3) and (4). In
 general, one trend prevails in each field but, in some fields, two or three
 trends appear, and this is indicated in column (2). In a few of fields, the
 star sample was poor, the fit was not possible and there is no evident trend.
 It is interesting to note that none one of the polarization trends is parallel
 to the mean direction of the Galactic Plane ($\theta$ $\sim$151$\degr$) toward
 this line of sight ({\it l}~$=$~256$\degr$, {\it b}~$=$~$-$9.2$\degr$). This
 suggests that the polarization data is indeed sampling the local magnetic field
 associated with the IVS region.  
 
To improve on the precision of the mean polarization values for each trend, we
 included a filter in each field. We select those objects with polarization
 angle between ($\theta_{gauss}$ $-$ 2$\times \sigma_{\theta gauss}$,
 $\theta_{gauss}$ $+$ 2$\times \sigma_{\theta gauss}$). The number ({\it N}) of
 stars with {\it P}/$\sigma_{P}$ $ \geq $ 10 in each filtered subsample is given
 in column (5). We then estimated the mean Stokes parameters,  
$\langle Q\rangle$ and $\langle U\rangle$, for each subsample, from the
 individuals values for each star ({\it q$_{i}$}, {\it u$_{i}$}) weighted by the
 error ($\sigma _{i}$) according to: 
 
\begin{eqnarray*} 
       \langle Q\rangle &   =     &  \sum ({\it q_{i}} / \sigma ^{2}_{i})   /
 \sum \sigma ^{-2}_{i}    \\ 
       \langle U\rangle &   =     &  \sum ({\it u_{i}} / \sigma ^{2}_{i})   /
 \sum \sigma ^{-2}_{i}.    \\ 
\end{eqnarray*} 
 
The estimated mean polarization value $\langle P\rangle$, its associated error 
 $\langle\sigma\rangle$ and its mean polarization position angle
 $\langle\theta\rangle$ are then given by: 
 
\begin{eqnarray*} 
 \langle P\rangle       &  =      & \sqrt{\langle Q\rangle^{2} + \langle
 U\rangle^{2}}         \\ 
 \langle\sigma\rangle   &  =      &       (1/\sum \sigma ^{-2}_{i})^{0.5}   \\ 
 \langle\theta\rangle & =       & 0.5 tan^{-1}(\langle U\rangle/\langle
 Q\rangle).         \\ 
\end{eqnarray*} 
 
These are presented in columns (6), (7) and (8). We show the mean error of the
 polarization angle in column (9). This was estimated from $\langle\sigma
 _{\theta}\rangle$ = $\sum \sigma _{\theta_{i}}/ N$ where
 $\sigma_{\theta_{i}}~=~28\fdg65\;\sigma _{{\it p_{i}}}/{\it p_{i}}$ is the
 individual error of the polarization position angle \citep{ser74} for an star
 with polarization ({\it p}$_{i}$  $ \pm $ $\sigma_{p_{i}}$). Column (10) shows
 the dispersion of polarization angles corrected in quadrature by its mean error
 ($\Delta \theta  =  \sqrt{\sigma_{gauss}^{2} -
 \langle\sigma_{\theta}\rangle^{2}}$); column (11) shows the squared of this
 dispersion (in radians).
 
\clearpage 
 
\begin{figure*} 
\includegraphics[scale=1]{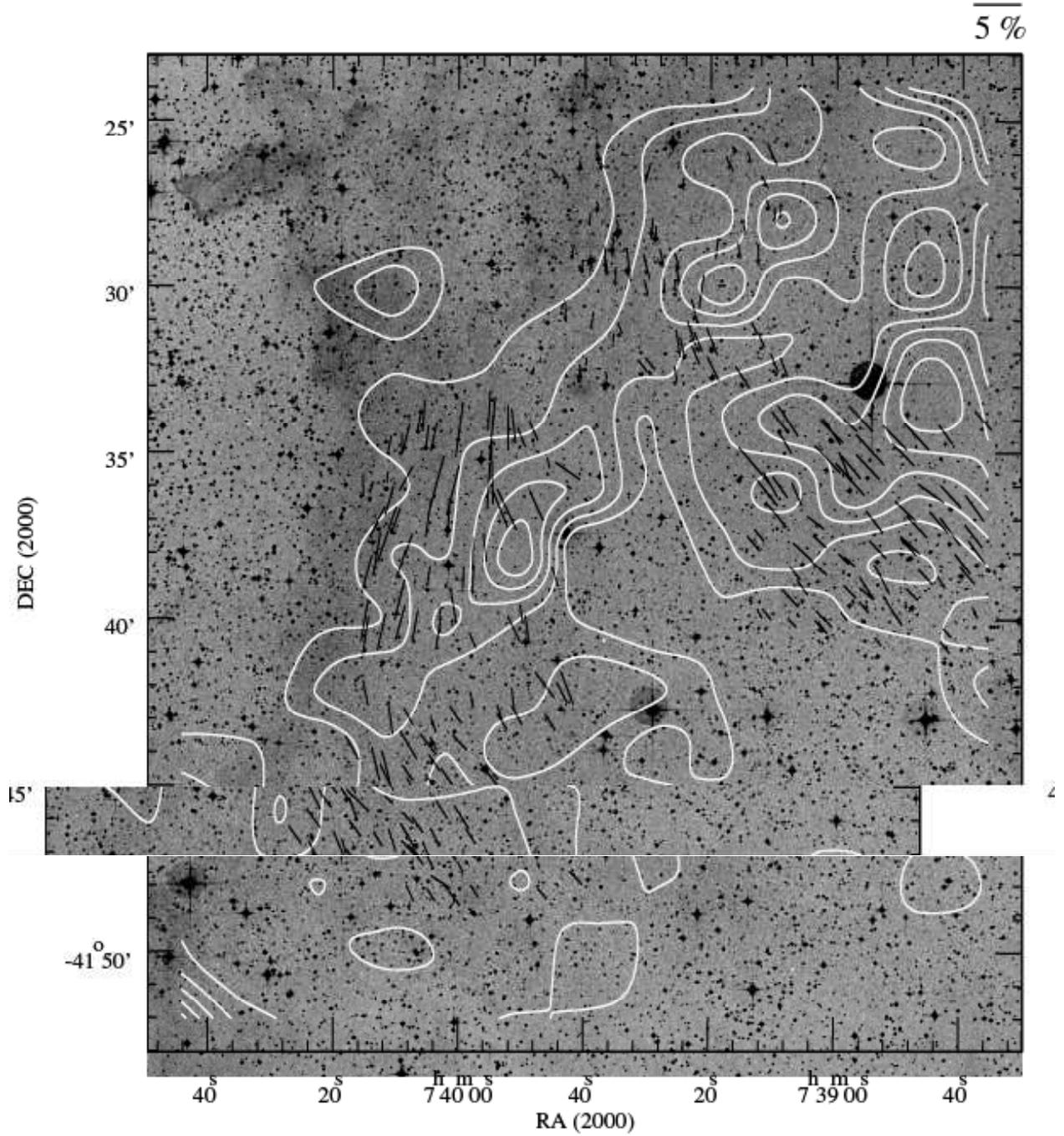} 
\caption{Polarization map overlapping the extinction map for fields 05, 06, 07
 and 08. The polarization scale is shown up to the right. The contours are the
 same as in Fig.~\ref{fcond}.\label{zoom1}} 
\end{figure*} 
 
 
 
\begin{figure*} 
\includegraphics[scale=1]{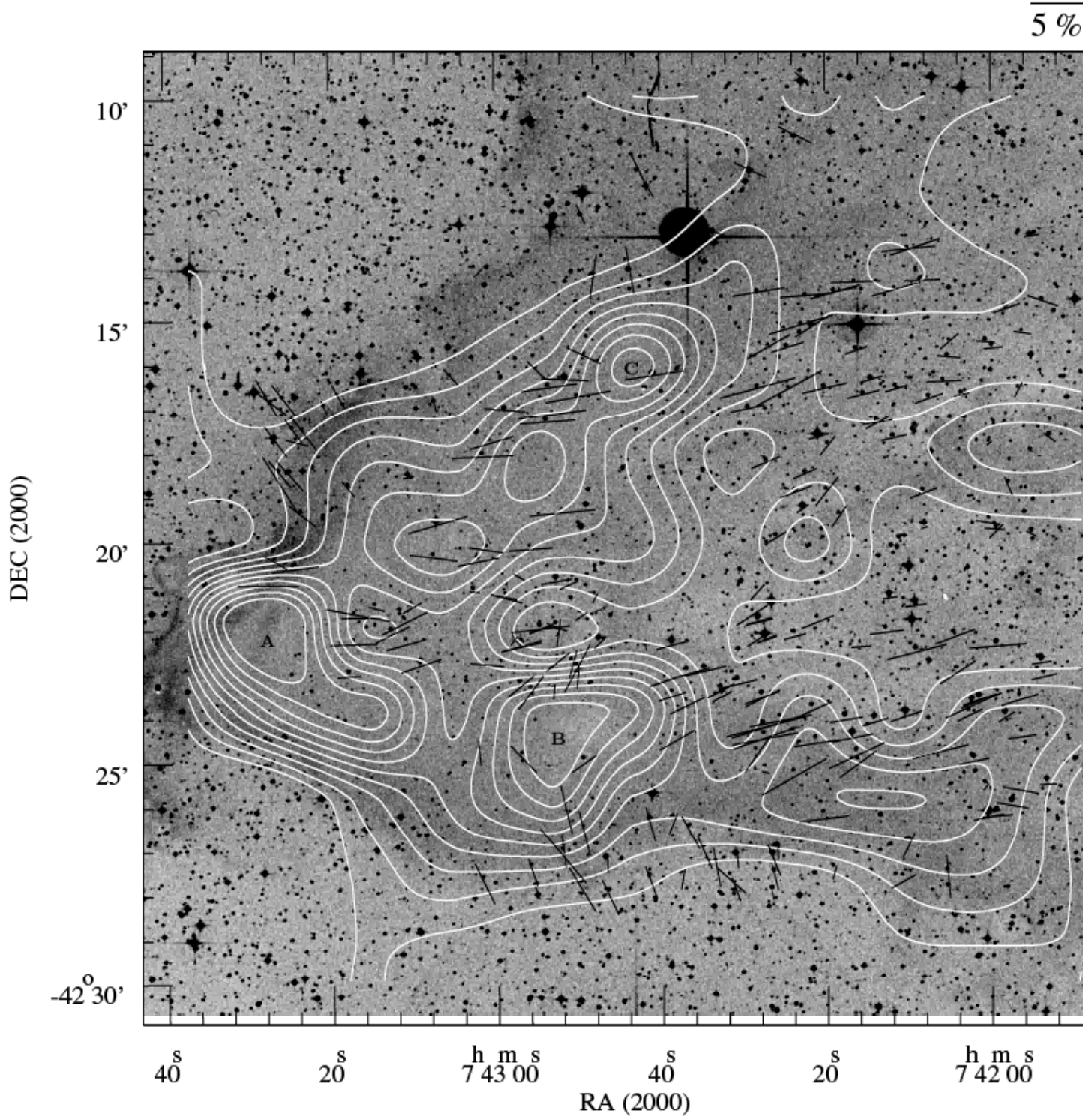} 
\caption{Polarization map overlapping the extinction map for fields 01, 04, 11,
 12 and 13. The polarization scale is shown up to the right. The contours are
 the same as in Fig.~\ref{fcond}.\label{zoom2} The condensations A, B and C are
 also indicated.} 
\end{figure*} 
 
\clearpage

\section{Polarization and extinction} 
\label{polext} 
 
To help our analysis of correlations between polarization and extinction, we
 estimated the average extinction for objects in each one of the polarization
 trends (column (12) of Table~\ref{tanl}). The position of each star in the
 polarimetric catalogue is well determined with a typical precision of less than
 1$\arcsec$. We then identified its position in the extinction maps and
 attributed the value of the extinction cell where the star was located. This
 procedure may underestimate the extinction value assigned to a star because the
 resolution of 2$\arcmin \times$2$\arcmin$ of the extinction maps may not
 resolve variations of extinction within one extinction cell.  
 
Figure~\ref{fextpol} plots the extinction map of Figure~\ref{fcond} along with
 the polarization map (Figure 5 in Paper I) of the region. The polarization
 vectors, which are from stars with ${\it P}/\sigma _{{\it P}}$ $ > $ 10 in the
 catalog, are concentrated in regions with significant extinction along the
 I-front. The extinction is low ({\it A}$_{\rm V}$ $ \leq $ 1.5 mag) but is
 enough to produce the observed polarization. The range of extinction sampled by
 our polarization data is similar to the extinction found in the outer layers of
 darks clouds and in the diffuse ISM \citep[{\it A}$_{\rm V}$ $<$ 3
 mag,][]{vrb85,ken94} and below the threshold extinction of the transition from
 bare to H$_{2}$O-mantled grains \citep[{\it A}$_{\rm V}$ $\sim$3.3
 mag,][]{ger95,whi01}.  
 
In order to quantify the interval of extinction sampled by our catalogue, we
 analysed two subregions (shown in Figures~\ref{zoom1} and~\ref{zoom2}) with
 well defined polarization patterns. 
 
Figure~\ref{zoom1} shows the four fields (05, 06, 07 and 08) at the top of
 northern ridge (see Figure~\ref{fhist}) around ($\alpha$,$\delta$)$_{2000}$ =
 (07$^{\rm h}$39$\fm$5,-41$\degr$37$\arcmin$). This region presents an observed
 polarization higher than average and a very smooth overall pattern as noted in
 Paper I. The range of mean polarization goes from 1.7\% to 2.5\% and the mean
 extinction from 0.32 to 0.45 mag. It is interesting to note that the fields 06
 and 07 (trend 1), closer to I-front wall and aligned with it, present a
 dispersion of polarization angle ($\Delta\theta$) at least twice than ones of
 the fields 05 and 08. It reinforces the suggestion of a magnetic field
 enhancement in the compressed postshock gas as indicated in Paper I. 
 
The region shown in Figure~\ref{zoom2} (fields 01, 04, 11, 12 and 13; see
 Figure~\ref{fhist}) presents a very complex polarization pattern. This
 complexity seems associated with the A, B and C extinction structures. This
 region has a range of mean polarization of 0.6\%$-$3.1\% with the highest value
 in field 13. The mean extinction goes from 0.33 to 0.47 mag. As we will see in
 \S\ref{polef}, the larger range of polarization values is a indication of
 significant variations in the polarization efficiency in small angular scales
 in this subregion. 
 
In some fields, our detection of trends through $\theta$ gaussian fits fails in
 cases where a good polarimetric sample seems to be present. This is evident in
 fields 09, 10 and 16 (see Figure~\ref{fhist}). Nevertheless, a non-detection of
 a polarization trend seems correlated with low extinction regions as can be
 verified by the mean extinction of fields 09 and 10 that are 0.21 mag and 0.19
 mag, respectively. However, field 16 has a high mean extinction (0.51 mag) but
 this value may be affected by the contrast gradient present at the border of
 the {\it DDS} image used.

\section{Magnetic fields} 
\label{mag} 
 
As we noted in \S\ref{intro}, CF gives an estimate of the magnetic field
 component ({\it B}) on the plane of the sky: 
 
\begin{equation} 
{\it B} = (4  \pi \rho)^{1/2} \nu   /   \Delta \theta   
\label{bdisp} 
\end{equation} 
 
where, {\it B} is in gauss (G), $\rho$ is the density (g cm$^{-3}$), $ \nu $ is
 the turbulent motion velocity (cm s$^{-1}$) and $ \Delta \theta $ is in
 radians. In terms of energy densities, the ratio of the kinetic energy density
 ($\rho _{kin}~\propto~\rho\nu^{2}$) to magnetic energy density 
($\rho_{mag}~\propto~B^{2}$) is the mean square fluctuation of  
{\it B} \citep{zwe90}, so 
 
\begin{eqnarray*} 
(\rho_{kin}/\rho_{mag}) \propto (\Delta\theta)^{2} .  
\end{eqnarray*} 
 
Therefore, $\rho $$_{kin}$/$ \rho $$_{mag}$ and/or {\it B} can be obtained from
 $ \Delta \theta $ (column (10) in Table~\ref{tanl}). In some fields (01, 09,
 10, 15, and 16), it was not possible to obtain information of $ \Delta \theta $
 because of a poor polarimetric sample or the polarimetric pattern is random. In
 the remaining fields, we estimated $(\Delta \theta)^{2}$ for each trend (column
 (11) of Table~\ref{tanl}). We can note that $(\Delta \theta)^{2}$ shows a range
 of (0.2$-$11.3)$\times$10$^{-2}$. This can be interpreted as variations of (at
 least) one order of magnitude in $\rho $$_{kin}$/$ \rho $$_{mag}$ along the
 region. Nevertheless, this large variation range observed in $(\Delta
 \theta)^{2}$ is consistent with the fact that some regions present random
 position angles. Extreme cases would be fields 09, 10 and 16, where the
 impossibility of estimating $\Delta \theta$ may be interpreted as the kinetic
 energy density prevailing over the magnetic energy density. 
 
\clearpage 
 
\begin{figure} 
\includegraphics[scale=1]{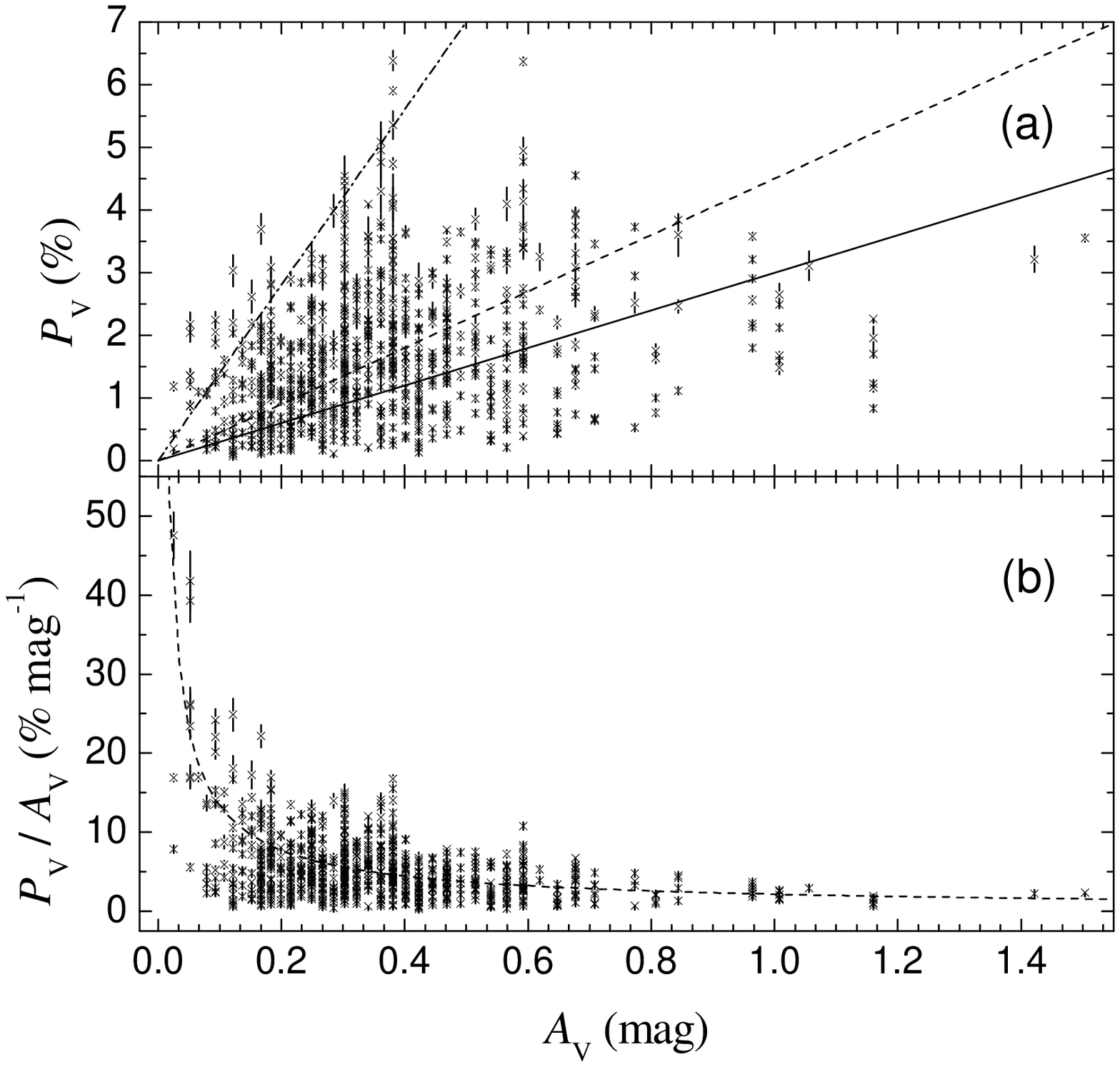} 
\caption{(a) Polarization versus visual extinction. The solid line represents
 the higher limit for ISM polarization \citep[{\it P}$_{max}$ $=$ 3{\it A}$_{\rm
 V}$;][]{ser75}. The dashed line represents the higher limit towards Cha I
 \citep[{\it P}$_{max}$ $=$ 4.5{\it A}$_{\rm V}$;][]{whi94}. The dot-dashed line
 is the upper limit ({\it P}~$=$~14{\it A}$_{\rm V}$) for dust grains consisting
 of completely aligned infinite dielectric cylinders \citep{whi92} that covers
 96.3$\%$ of the sample. (b) Polarizing efficiency versus extinction. The dashed
 line represents a least-squares power-law fit to our sample ({\it P}$_{\rm
 V}$/{\it A}$_{\rm V}$~=~2.54$\pm$0.08~{\it A}$_{\rm
 V}$$^{-0.61\pm0.04}$~\%~mag$^{-1}$). \label{fepl}} 
\end{figure} 
 
 
 
\begin{figure} 
\includegraphics[scale=1]{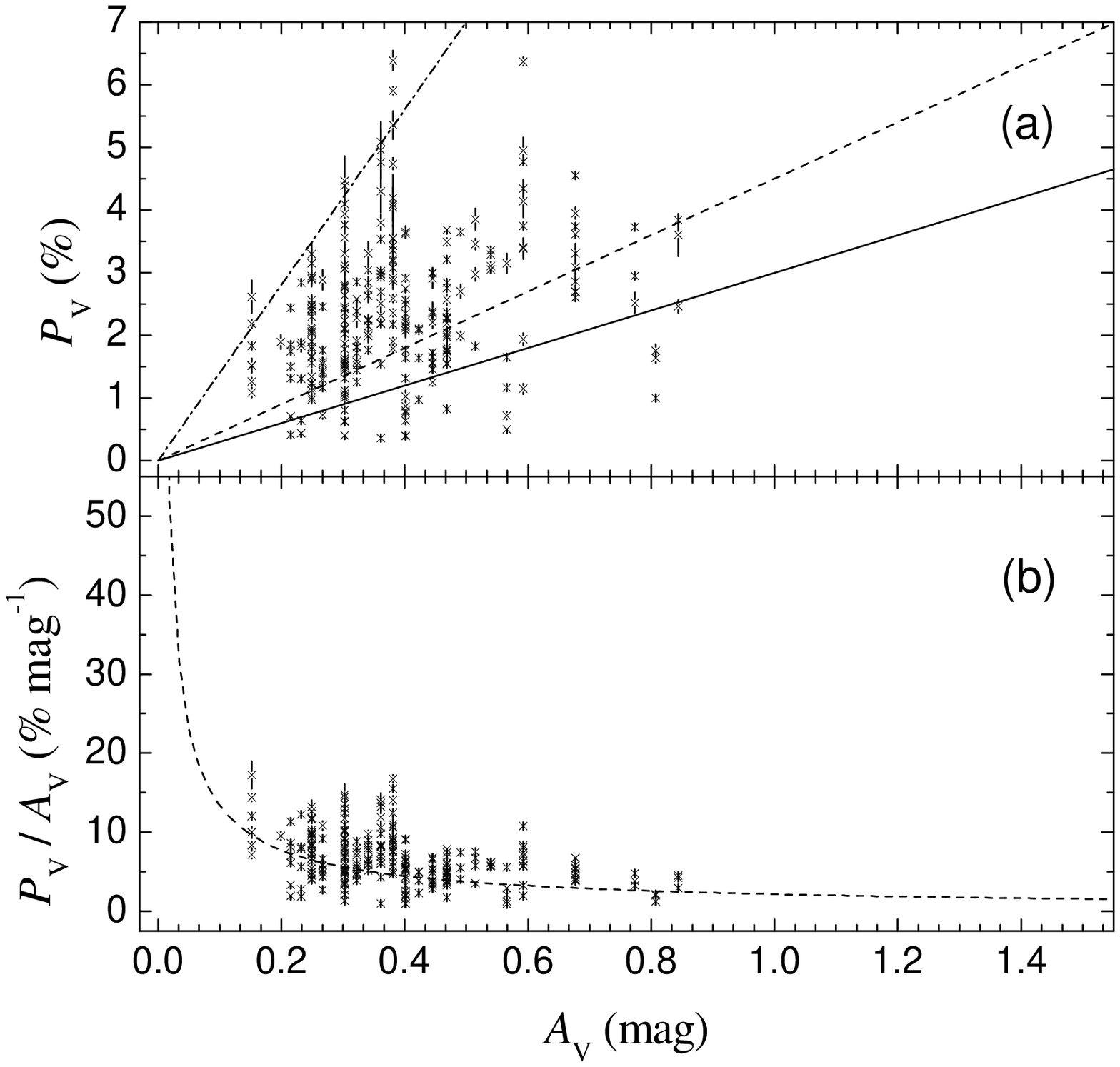} 
\caption{(a) Polarization versus visual extinction for fields in
 Figure~\ref{zoom1}. The solid, dashed and dot-dashed lines as in
 Figure~\ref{fepl}. (b) Polarizing efficiency versus extinction. The dashed line
 represents a least-squares power-law fit to this sub-sample ({\it P}$_{\rm
 V}$/{\it A}$_{\rm V}$~=~3.50$\pm$0.25~{\it A}$_{\rm
 V}$$^{-0.60\pm0.07}$~\%~mag$^{-1}$). \label{zoom1epl}} 
\end{figure} 
 
 
 
\begin{figure} 
\includegraphics[scale=1]{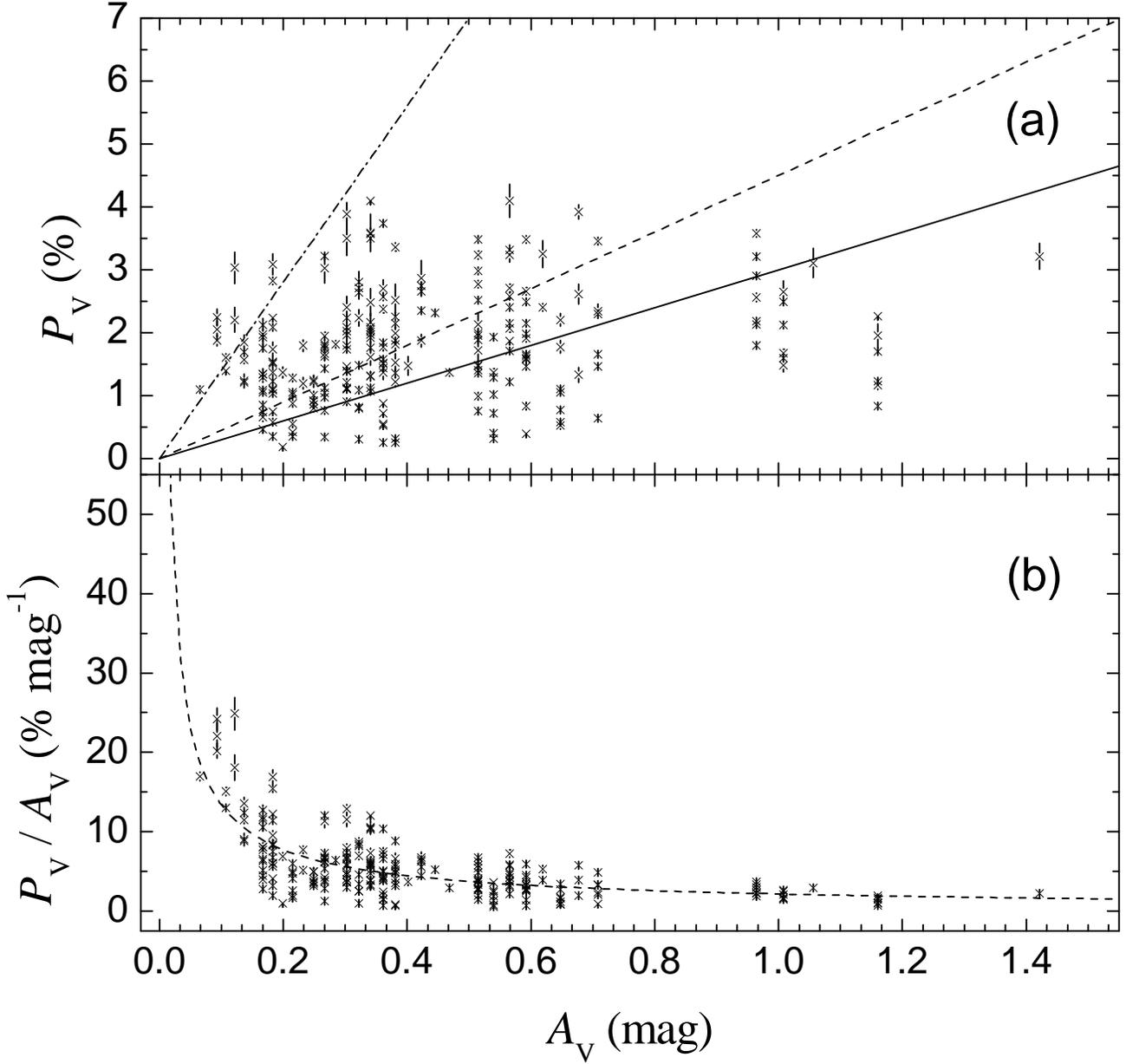} 
\caption{(a) Polarization versus visual extinction for fields in
 Figure~\ref{zoom2}. The solid, dashed and dot-dashed lines as in
 Figure~\ref{fepl}. (b) Polarizing efficiency versus extinction. The dashed line
 represents a least-squares power-law fit to this sub-sample ({\it P}$_{\rm
 V}$/{\it A}$_{\rm V}$~=~2.16$\pm$0.14~{\it A}$_{\rm
 V}$$^{-0.79\pm0.07}$~\%~mag$^{-1}$). \label{zoom2epl}} 
\end{figure} 
 
\clearpage 
In order to calculate the strength of the magnetic field, we assume that the
turbulent velocity ($\nu$) is represented by the dispersion velocity of a
Gaussian distribution ($\Delta V$$^{2}$~=~8ln2 $\nu$$^{2}$) where $\Delta V$
is obtained from the mean FWHM of an appropriate molecular line. This dispersion
may represent the gas, of density $\rho$, in which the polarization arises.
The CS emission in this region yields $\Delta V_{\rm CS}$~=~1.4 km s$^{-1}$ \citep{chu96}. 
As noted by these authors, the density, $\rho~=~{\it mn}(\rm H_{2})$, needed for 
CS excitation requires a particle density {\it  n}(H$_{2}$) $\sim$10$^{4}$cm$^{-3}$.  
As the CS emission samples the higher density cores, its density may represent
a higher limit for the outer layers of dust clouds which our polarization 
measurements sample. \cite{vil00} analised CO emission in the Vela region and 
found mean dispersions of $\Delta V_{\rm ^{13}CO}$~=~0.7$\pm$0.2~km s$^{-1}$ and 
$\Delta V_{\rm C^{18}O}$~=~0.6$\pm$0.2~km s$^{-1}$. As the CO emission probably
samples the envelope regions of molecular clouds, its dispersion may better 
represent the turbulent motions. Therefore, we believe that a dispersion of
$\Delta V$~=~1.0~km s$^{-1}$ is a good compromise to consider as input parameter
in our computation.

We used the average extinction (column (12) of Table~\ref{tanl}) to infer the mean particle density $\langle\it n\rm (H_{2})\rangle$ in each polarization trend. We take,

\begin{eqnarray*}
 \langle\it n\rm (H_{2})\rangle & = & \langle A_{\rm V}\rangle  ({\it N}_{\rm H_{2}}/{\it A}_{\rm V}) \frac{1}{{\it l}} , \\
\end{eqnarray*}

where {\it l} is the typical size (in parsecs) of an extinction cell assuming a distance of 450 pc. The computed 
values are indicated in column (13) of Table~\ref{tanl}. The overall average density from the trends is 
312$\pm$97 cm$^{-3}$.  This value is consistent with our sampling the low density regions, as discussed above, and will be used in the following. 

Therefore, with the above parameters and using eq.~\ref{bdisp}, the magnetic field
 expression becomes: 
 
\begin{eqnarray*} 
     {\it B}(\rm G) & = &  2.88 \times 10^{-6} (\frac{{\it n}(\rm H_{2})}{312
 {\;\rm cm}^{-3}})^{1/2}(\frac{\Delta V}{1.0 {\;\rm km \;s}^{-1}}) /  \Delta
 \theta .  \\ 
\end{eqnarray*} 
 
In this way, we estimated {\it B} (column (14) of Table~\ref{tanl}) for each
 field and for each trend with a well defined dispersion of polarization angle
 ($\Delta \theta$). \citet{ost01} noted that $\Delta\theta$ $<$ 25$\degr$ yields
 a good estimate of the plane-of-sky magnetic field strength if a multiplying
 factor of $\sim$0.5 is applied to CF formula. As we can note in
 Table~\ref{tanl}, all the $\Delta\theta$ values are lower than this limit and
 the above expression also include this correction. As usual, use of the CF 
formula for estimating {\it B} carries an uncertainty from all discussed here. 
Apart from the above mentioned multiplying factor, the uncertainties in 
${\it n}(\rm H_{2})$, $\Delta V$ and $\Delta\theta$ imply a typical uncertainty 
of at least 50\% in {\it B}.
 
The average strength of the magnetic field along the I-front is
 [0.018~$\pm$~0.013] mG and the full range of variation is $\sim$one order of
 magnitude (0.007$-$0.064 mG). As \citet{zwe90} pointed out, if the line of sight
 samples $\mathcal{N}$ different regions with different magnetic orientations,
 {\it B} would tend to be overestimated by $\sqrt{\mathcal{N}}$. This averaging
 effect may be important, for example, in the region of complex polarization
 pattern showed in Figure~\ref{zoom2} where {\it clumpy} structures are present.

Finally, \cite{chu96} concluded that the potential energy of this region is just
 $\sim$0.05 of the total kinetic energy (thermal plus turbulent) and the
 expanding shell pressure avoids the dissipation of the region. From the
 estimates of {\it B} along with the observed fact that, in general, the
 magnetic pressure dominates over the turbulent pressure in several parts of the
 shell, we conclude that the magnetic component also contributes in an important
 way to the dynamical balance of this region. 
 
\section{Polarizing efficiency} 
\label{polef} 
 
We combined our polarization data with the extinction maps obtained by star
 counts to analyze the polarizing efficiency of the dust in the region.
 Figure~\ref{fepl}a plots the visual polarization percentage versus the visual
 extinction for the 856 objects in the catalogue.  
 
As we noted above, the extinction range in our sample extends up to $\sim$1.5
 mag and the maximum polarization is $\sim$6\%. The solid line in
 Figure~\ref{fepl}a represents the upper limit for optimum polarization
 efficiency in diffuse ISM \citep[{\it P}$_{max}$ $=$ 3{\it A}$_{\rm
 V}$;][]{ser75} and just 30.2\% of the sample is under this limit. \citet{whi94}
 found a higher upper limit for optimum polarization ({\it P}$_{max}$ $=$
 4.5{\it A}$_{\rm V}$) toward some lines of sight in the Cha I dark cloud. They
 concluded that a high degree of alignment is present with the magnetic field
 lines essentially perpendicular to our line of sight. If we consider this upper
 limit  (dashed line in Figure~\ref{fepl}a) the percentage of our sample under
 this limit rises to 49.8\%.  
 
It seems to be clear that part of the dust that produces the observed alignment
 in this region has different properties than diffuse ISM dust and it is
 especially true for low extinction ({\it A}$_{\rm V}$ $<$ 0.8 mag) regions. On
 the other hand, the higher polarization efficiency for low extinction can be a
 indicator that an optimum alignment is present in some regions of I-front maybe
 favored by a privileged geometric view (i.e. parallel to plane of the sky) of
 the magnetic field with respect to the wall seen edge-on. This fact can be
 quantified by the theoretical upper limit (${\it P}_{\rm V}/{\it A}_{\rm V} \la
 14$) for dust grains consisting of completely aligned infinite dielectric
 cylinders \citep{whi92} that covers 96.3\% of the sample (dot-dashed line in
 Figure~\ref{fepl}a).   
 
Nevertheless, and as mentioned in \S\ref{polext}, the possibility of our
 2$\arcmin \times$2$\arcmin$ extinction cell will be underestimating the
 extinction can not be discarded. If this is the case, higher extinction values
 can be masked by the observed clumpy structures and an optimum alignment better
 than different dust properties is more plausible to explain the correlations
 found. 
 
\clearpage 
 
\begin{figure} 
\plotone{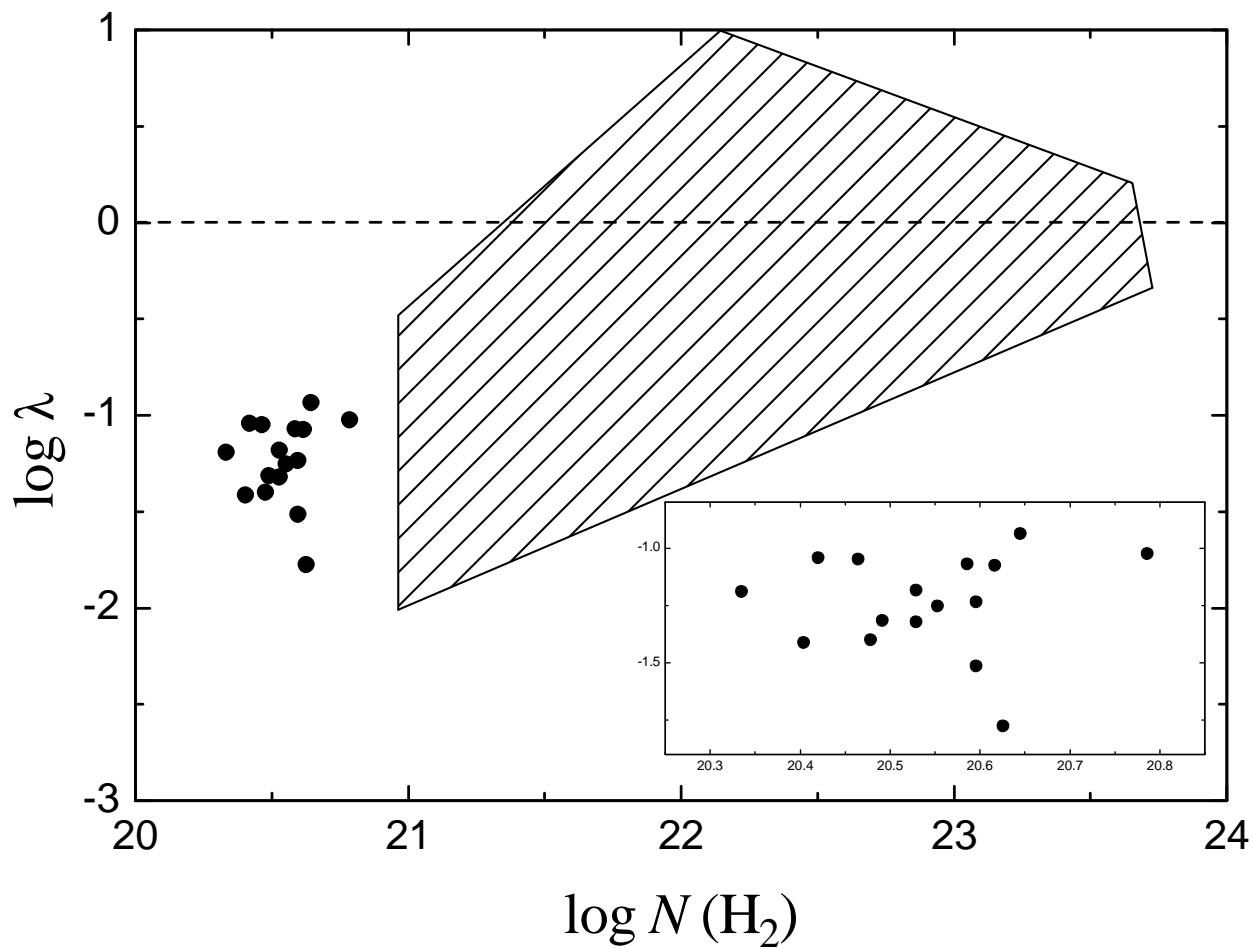} 
\caption{Observed mass-to-magnetic flux ratio in units of the critical value
 (corrected by bias projection) versus molecular column density. In black dots
 are the $\lambda$s evaluated in each polarization trend (with {\it B}
 estimated) from Table \ref{tanl}. The IVS region is dominated by the
 subcritical regime ($\lambda < 1$). The dashed region represents all the
 $\lambda$s available for molecular clouds \citep{crut04}. The box inside the
 figure is a zoom to show more detail.  \label{lambda}} 
\end{figure} 
 
\clearpage 
 
In Figure~\ref{fepl}b, we plot the polarizing efficiency versus the extinction
 and it is very clear as regions with low extinction are more efficient to
 polarize the light of background stars than high extinction regions. This fact
 already has been observed previously in darks clouds
 \citep{vrb76,vrb81,vrb92,vrb93,mcg94,whi94,ger95,god95}. However, in I-fronts
 of shells in expansion, as this study, previous evidence does not exist in
 literature. 
 
The dashed line in Figure~\ref{fepl}b shows a least-squares power-law fit to our
 sample ({\it P}$_{\rm V}$/{\it A}$_{\rm V}$ = 2.54$\pm$0.08~{\it A}$_{\rm
 V}$$^{-0.61\pm0.04}$~\%~mag$^{-1}$). The dependence of the polarizing
 efficiency with the extinction found by us is very close to the models of
 \citet[{\it P}/{\it A}~$\propto$~{\it A}$^{-0.5}$]{jon89} that assumes a
 magnetic field with random and uniform components. Our fit also compares very
 well with \citet[{\it P}/{\it A}~$\propto$~{\it A}$^{-0.56\pm0.17}$]{ger95} for
 observed data in Taurus dark cloud.  
 
The ratio {\it P}/{\it A} may be written as {\it P}/{\it A}~$\propto$~{\it
 A}$^{2{\it k}-1}$ \citep{ger95}, where {\it k} represents the amplification of
 the magnetic field with a gas density \citep[{\it B}~$\propto$~{\it
 n}$^{k}$]{mou78}. Our fit yields {\it k}~=~0.195. This value is below the
 observed lower limit ({\it k}~=~0.33) of \citet{mou78} and the models of
 magnetic fields in turbulent molecular clouds \citep{ost01}. As pointed out by
 \citet{ger95} this discrepancy could be result of ignoring effects of
 small-scales inhomogenties in the magnetic field and of the coupling of gas and
 dust temperatures in regions of high density in the assumed polarizing
 efficiency \citep[{\it P}/{\it A}~$\propto$~{\it B}$^{2}$/{\it n}
 from][]{vrb81,vrb93}. Finally, the coefficient of our fit (2.54) is close to
 {\it P}/{\it A}~=~3 \citep[from][]{ser75} and this suggests that the higher
 extinction regions (0.8~$<$~{\it A}$_{\rm V}$~$<$~1.5) of our sample are more
 {\it normal} and comparable to diffuse ISM.

As comparison, we made the same as analysis for the two subsamples shown in
 Figures~\ref{zoom1} and~\ref{zoom2}. The results are shown in
 Figures~\ref{zoom1epl} and~\ref{zoom2epl}, respectively. As we can note, the
 region sampled in Figure~\ref{zoom1} corresponds to low extinctions, below
 $\sim$0.8 mag. We find a high polarizing efficiency and this subsample seems to
 be representative of the overall trend observed in the full sample
 (Figure~\ref{fepl}). Note that the exponents of the fits for this subsample and
 the full sample are essentially the same.  On the other hand, the very complex
 polarization pattern shown in the region indicated in Figure~\ref{zoom2} has a
 higher extinction range (up to $\sim$1.5 mag, Figure~\ref{zoom2epl}) and the
 polarizing efficiency is lower if compared with the previous region.

\section{Mass-to-flux ratio} 
\label{masflux} 
 
\citet{car88} noted that {\it IRAS} point sources are coincident with the
 densest portion of this region. They speculated that shocks might have
 triggered local collapse and possibly star formation. However, they associated
 the clumpy structure seen in this region with clumps of warm dust with peak
 emission in 60$\mu$m band and star formation was ruled out.  
 
The importance of magnetic fields in the evolution of interstellar clouds and
 star formation can be tested if the mass-to-magnetic flux ratio can be
 quantified ({\it M}/$ \Phi $). We can evaluate this issue using the $\lambda$
 parameter \citep[][and references therein]{cru04} defined as:

\begin{equation} 
\lambda \equiv \frac{(M /\Phi)_{actual}}{(M /\Phi)_{crit}} = 7.6 \times 10^{-21}
 \frac{{\it N}(\rm H_{2})}{\it B},  
\label{eqlambda} 
\end{equation}

where ({\it M}/$\Phi$)$_{crit}$ is the critical value for the mass that can be
 supported by a magnetic flux, {\it N}(\rm H$_{2}$) is in cm$^{-2}$ and {\it B}
 is in $\mu$G. In the subcritical regime ($\lambda < 1$), magnetic support
 prevents the collapse and the star formation and probably this is the case in
 the IVS region.

We evaluated the $\lambda$ parameter (using eq.~\ref{eqlambda}) for each
 polarization trend where {\it B} has been estimated. The results are indicated
 in the last column of Table~\ref{tanl}. We obtained {\it N}(H$_{2}$) from the
 mean extinction $\langle A_{\rm V}\rangle$ for each trend (column (12) in
 Table~\ref{tanl}) and using the standard gas-to-extinction ratio (as in
 \S\ref{extmas}). As $\langle A_{\rm V}\rangle$ and {\it B} are estimated in
 exactly the same area, this guarantees a proper determination of $\lambda$.
 Clearly, all the regions with an alignment trend are in the subcritical regime
 ($\lambda <$ 1) between 0.05 and 0.35. If we applied the 1/3 correction factor
 to take in count the projection bias \citep{cru04}, the subcritical regime is
 even sharper. This is evident in Figure~\ref{lambda} that shows the
 distribution of $\lambda$ (corrected by projection bias) as a function of the
 molecular column density.   
 
It seems clear that magnetic support is important in the IVS region and the data
 are consistent with the prediction of ambipolar diffusion models with cloud
 envelopes being initially subcritical. This is also consistent with non-star
 formation in this region as speculated by \citet{car88}. 
 
It is interesting to note that the two points with the largest column densities
 and largest $\lambda$s (Figure ~\ref{lambda}) correspond to polarization trends
 associated with the high extinction structures A and B (see \S\ref{extmas}).
 \citet{crut04} analyzed all the $\lambda$s available for molecular clouds
 (dashed region in Figure~\ref{lambda}) and noted a slight indication that for
 large colum densities, $\lambda$ may be supercritical, and for small column
 densities, subcritical. Alhough that our data in Figure~\ref{lambda} samples
 just a interval of lower {\it N}(H$_{2}$) values ($\sim$10$^{20}$ cm$^{-2}$),
 evidence of $\lambda$ rises with {\it N}(H$_{2}$) can not be discarded. 
 
\section{Conclusions} 
\label{concl} 
 
Our conclusions can be summarized as follows: 
 
1. Extinction maps obtained using automatic star counts toward the western side
 of {\it IRAS} Vela Shell yield evidence of clumpy structures with typical lower
 limits of masses between 1.5~and~6~{\it M}$_{\sun}$ and a length scale {\it
 L}~$\sim$0.47~pc (assumed a distance of 450~pc).  
 
2. We estimated the strength of the local magnetic field on the plane of the sky
 using the dispersion of polarization angles and educated estimates for other cloud properties. 
The mean value computed along the  I-front was [0.018$\pm$0.013] mG and the full range of variation was
 (0.007$-$0.064)~mG. In terms of energy densities, the magnetic pressure generally
 dominates over the turbulent motions along the I-front. In a few cases, the
 polarization angle appears to change randomly, suggesting that the kinetic
 energy can dominate. 

3. We also investigated the polarizing efficiency combining the polarization
 data with the extinction maps. We found high polarizing efficiency in low
 extinction regions. It can be explained by different properties of dust as
 compared with the general ISM dust. Another more plausible possibility to
 explain this high polarizing efficiency is an optimum mechanism of dust
 alignment maybe favored by the magnetic field lines are nearly parallel to the
 plane of the sky and perpendicular to the I-front. 
 
4. Using the strength of magnetic field and the extinction estimated in each
 polarization trend was possible to evaluate the mass-to-magnetic flux ratio
 along the IVS region. The range of the $\lambda$ parameter we found
 (0.05$-$0.35) confirms that the magnetic support is dominant along this I-front
 and a subcritical regime prevails. The $\lambda$$-${\it N}(H$_{2}$) relation we
 find joins smoothly the overall relation previously obtained for higher density
 regions. This provides general support for the evolution of initially
 subcritical clouds to an eventual supercritical stage. 
 
\acknowledgments 
 
The authors wish to thank the anonymous referee for his/her careful reading and suggestions that helped to improve the paper. AP is thankful to CAPES and FAPESP (grant 02/12880-0) for financial support. AMM
 acknowledges support from FAPESP and CNPq. Polarimetry at IAG-USP is supported
 by a FAPESP grant 01/12589-1.

\clearpage 
 
\newpage

\clearpage 
\end{document}